\documentclass{article}%
\usepackage{amsfonts}
\usepackage{amsmath}
\usepackage{amssymb}
\usepackage{graphicx}%
\providecommand{\U}[1]{\protect\rule{.1in}{.1in}}

\begin{document}

\title{Instability and quantization in quantum hydrodynamics}
\author{Yakir Aharonov$^{\dag\ast}$ and Tomer Shushi$^{\ast\ast}$\\$^{\dag}$Schmid College of Science, Chapman University, Orange\\CA 92866, USA\\$^{\ast}$Raymond and Beverly Sackler School of Physics\\and Astronomy Tel-Aviv University, Tel-Aviv\\69978, Israel\\$^{\ast\ast}$Center for Quantum Science and Technology\\\& Department of Business Administration,\\Guilford Glazer Faculty of Business and Management,\\Ben-Gurion University of the Negev, Beer-Sheva, Israel}
\maketitle

\begin{abstract}
In this short paper, we show how a quantum nonlocal effect of far-apart
wavepackets in the Schr\"{o}dinger picture of wavefunctions is replaced by a
local instability problem when considering\ the hydrodynamical formulation of
quantum mechanics, known as the Madelung picture.\ As a second result, we show
how the Madelung equations describe quantized energies\ without any external
quantization conditions.


\end{abstract}

\section{\bigskip Introduction}

Nonlocality is one of the cornerstones of quantum mechanics and plays a key
role in the theory of quantum interference (see, e.g., [1]). Over the years,
several theories have been proposed to provide an alternative interpretation
of quantum mechanics, which provides a different story about the nature of the
particles. We are going to focus on one of these alternatives, namely, the
theory of quantum hydrodynamics (QH) [2-7]. In this report, we show how one
can reformulate the problem of nonlocal behavior of the phase difference of a
particle described by far-apart wavepackets\ into a local problem with
instability as a direct consequence of such conversion from nonlocality to
locality. The Madelung equations reformulate the Schr\"{o}dinger equation into
classical equations of motion, with the addition of the quantum potential.
Following the Schr\"{o}dinger equation with some Hamiltonian $H,$
$i\partial_{t}\Psi=H\Psi,$ by taking the polar representation of the
wavefunction $\Psi=Re^{iS},$ we obtain a system of coupled equations with
observables $\left(  \rho,J\right)  $ where $\rho=R^{2}$ is the density
function and $J:=\frac{\partial}{\partial x}S\cdot\rho$ being the current
density, subject to some initial datum $\left(  \rho\left(  x,0\right)
,J\left(  x,0\right)  \right)  \in%
\mathbb{R}
_{+}\times%
\mathbb{R}
.$ Since both $\rho$ and $J$ are gauge-invariant, there is no need to consider
potentials, which play an essential role in the linear differential equations.
Following the Madelung equations, the density function and the current density
are considered physical quantities of the hydrodynamical description of the
quantum particle. The Madelung formulation of quantum mechanics allows us to
get a new perspective in understanding quantum mechanics and also its relation
with classical mechanics since the Madelung equations are based on a
continuity equation of a fluid and a Hamilton-Jacobi equation of such a fluid,
with the addition of the quantum potential (see, [8-12]). In the following, we
show new basic properties and behavior of the Madelung equations that touch on
the foundations of the hydrodynamical formalism of quantum mechanics.

\section{Instability in the Madelung formalism}

Suppose that we have a superposition of two far-apart spatially separated
narrow Gaussian wavepackets with width $\Delta x<<L$ positioned at $-L$ and
$L,$ $L>>0,$ where the wavepacket located in $x=L$ move towards the other
wavepacket with momentum $p_{0}\,$, such that at the region $[-\ell,\ell],$
$0<\ell<L,$ the density function is significantly small$\ \rho\left(
x\right)  \approx\varepsilon^{N},$ $N>1,$ such that $\varepsilon<<\ell/\Delta
x.$ In the Schr\"{o}dinger picture, the\ wavefunction is given by%
\begin{equation}
\Psi\left(  x,t=0\right)  =\frac{1}{\sqrt{2}}\phi\left(  \frac{x+L}{2}\right)
+\frac{1}{\sqrt{2}}e^{ip_{0}x/\hbar}\phi\left(  \frac{x-L}{2}\right)  ,
\label{Gauss1}%
\end{equation}
where $\phi\left(  u\right)  =e^{-u^{2}/2\sigma^{2}}/\sqrt{2\pi\sigma
}\mathcal{N}$ is a Gaussian function with dispersion parameter $\sigma>0,$ and
$\mathcal{N}>0$ is the normalizing constant of $\Psi$.

\includegraphics[scale = 0.4]{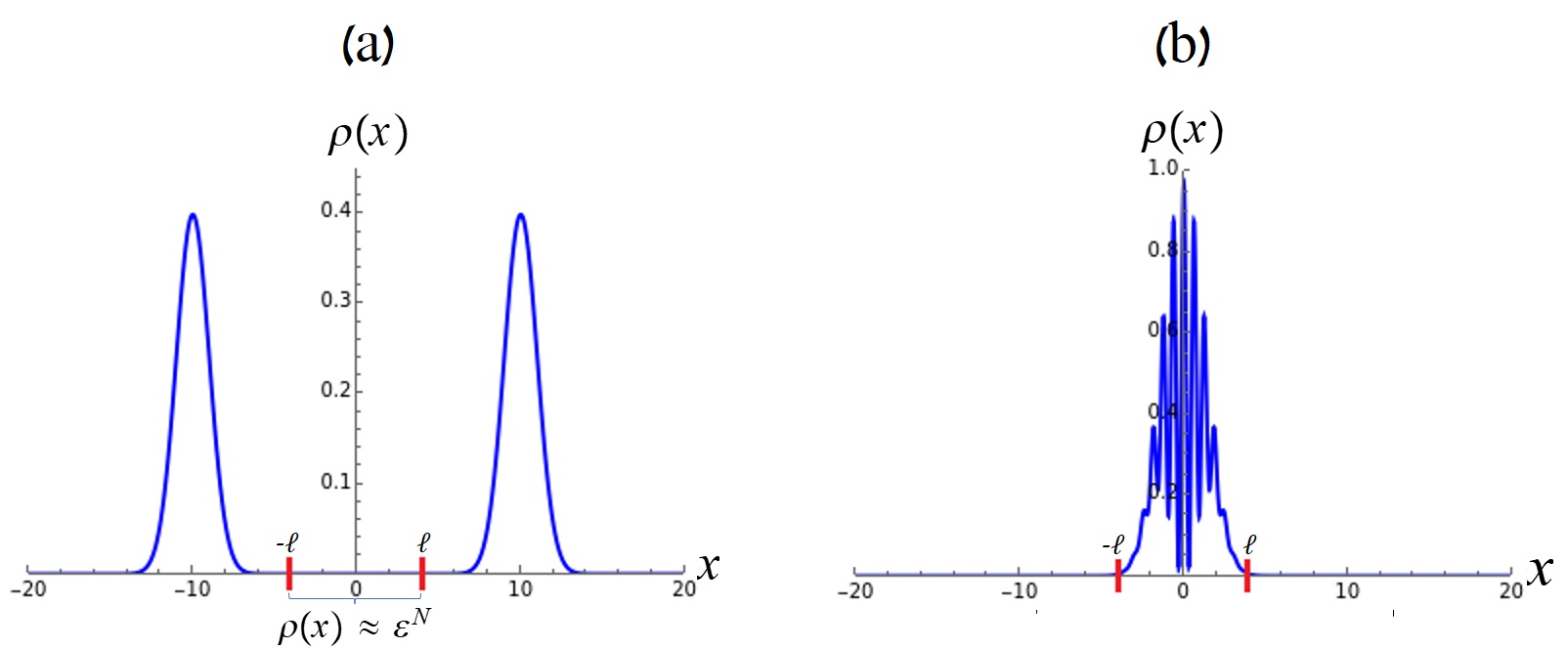}

Figure 1. The pair of Gaussian wavepackets before (a) and after (b) their interference.

\bigskip

As illustrated in Figure 1, the initial state shows that the far-apart
(Gaussian) wavepackets do not reveal their relative phase, and so any
measurement of one of the wavepackets will not reveal the relative phase. Only
after the wavepackets approach close enough to each other we have an
interference pattern. This basic quantum effect has a profoundly different
essence when dealing with the hydrodynamical description of quantum mechanics,
as will soon be discussed.

Let us now observe the effect in quantum hydrodynamics when considering
$\left(  \rho,J\right)  $ as the description of the quantum system in the form
of a fluid.

\bigskip At the region $[-\ell,\ell],$ the\ phase difference is given by the
integral of $J/\rho$ at that region,%
\begin{equation}
S^{\ell}\left(  t\right)  =\int_{-\ell}^{\ell}\frac{J\left(  x,t\right)
}{\rho\left(  x,t\right)  }dx. \label{3}%
\end{equation}
Suppose now that we have a small change $\varepsilon$ in the initial condition
of $J\left(  x,0\right)  $. Then, the phase difference at the interval
$[-\ell,\ell]$ is given by%
\begin{equation}
S_{+\varepsilon}^{\ell}\left(  t=0\right)  =\int_{-\ell}^{\ell}\frac{J\left(
x,0\right)  +\varepsilon}{\rho\left(  x,0\right)  }dx=S^{\ell}\left(
t=0\right)  +\varepsilon\int_{-\ell}^{\ell}\frac{1}{\rho\left(  x,0\right)
}dx \label{4}%
\end{equation}
so for $\rho\left(  x\right)  \approx\varepsilon^{N}$ at the region
$[-\ell,\ell],$ we have%
\begin{equation}
S_{+\varepsilon}^{\ell}\left(  t=0\right)  \approx S^{\ell}\left(  t=0\right)
+\frac{2\ell}{\varepsilon^{N-1}}, \label{5}%
\end{equation}
with $2\ell/\varepsilon^{N-1}>>1.$ As an illustrative example of the effect,
in the case of (\ref{Gauss1}), we have%
\begin{align}
S_{+\varepsilon}^{\ell}\left(  t=0\right)   &  \approx\arctan\left(
\frac{\sin\left(  p_{0}x/\hbar\right)  }{\phi\left(  \frac{x+L}{2}\right)
/\phi\left(  \frac{x-L}{2}\right)  +\cos\left(  p_{0}x/\hbar\right)  }\right)
\left(  |_{x=\ell}-|_{x=-\ell}\right)  +\frac{2\ell}{\varepsilon^{N-1}%
}\nonumber\\
&  \approx\frac{2\ell}{\varepsilon^{N-1}}. \label{SSS11}%
\end{align}

The instability effect does not depend on the shape of the wavefunction as
long as the given conditions are fulfilled, e.g., having separated wavepackets
with $\rho\left(  x\right)  \approx\varepsilon^{N}$ at the region $[-\ell
,\ell]$ for some arbitrary small $\varepsilon$ and some integer $N>1.$
Moreover, the growth rate of the initial phase difference is also the same for
every shape of the wavefunction (as long as the wavefunction does not depend
on $\varepsilon$ explicitely). In general, we have%
\begin{equation}
S_{+\varepsilon}^{\ell}\left(  t=0\right)  \approx\frac{1}{\varepsilon^{N}%
}\int_{-\ell}^{\ell}J\left(  x,0\right)  dx+\frac{2\ell}{\varepsilon^{N-1}}.
\label{S_Plus}%
\end{equation}

The instability effect can be seen as an \textit{amplification effect} where
small changes in the environment of the wavepackets will be revealed as a
large change in their observed phase difference. We identified that when the
local formulation of quantum mechanics\ is considered, there is a natural
occurrence of instability in the relation between the current density and the
density function, causing a dramatic change when the initial current density
possesses a significantly small change $\varepsilon.$ This instability seems
to be a\ compensation that comes with the\ local formulation, which is
switched off when dealing with nonlocal pictures such as the Schr\"{o}dinger
or the Heisenberg formulations of quantum mechanics.

For future research, we suggest looking at more complex quantum systems and
exploring them following quantum hydrodynamics with finding traces of
instabilities that emerged from such a formulation.

\section{Madelung equations without quantization conditions}

Soon after the Schr\"{o}dinger equation was introduced to the world by Erwin
Schr\"{o}dinger, the hydrodynamic formulation of quantum mechanics was
proposed by Erwin Madelung. A fundamental question is whether Madelung's
equations genuinely describe the same physics as in the Schr\"{o}dinger
equation or not.

We show how the Madelung equations indeed describe quantized physics without
any quantization conditions. In particular, we show that when quantized energy
levels are solutions of the Schr\"{o}dinger equations, they are also solutions
of the Madelung one -- they are a complete set of solutions. We start by
observing that the Madelung equations consider two physical quantities, $\rho$
and $J,$ instead of one, the wavefunction. We recall that $\rho$ describes the
density function, and thus, it is assumed to be integrable, normalized, and
satisfy
\begin{equation}
\rho\left(  x\right)  \rightarrow0\text{ \ \ \ as \ \ }\left\vert x\right\vert
\rightarrow+\infty. \label{rho_1}%
\end{equation}

We now show how the property (\ref{rho_1}) imposes quantized energy levels of
the Schr\"{o}dinger equation. First, we consider some potential $V\left(
x\right)  ,$ with the spatial part of the Schr\"{o}dinger equation
$-\frac{d^{2}\psi}{dx^{2}}+V\left(  x\right)  \psi=E\psi.$ Following the
Madelung representation and taking the real part, we have%
\begin{equation}
-R^{\prime\prime}\left(  x\right)  +S^{\prime2}R\left(  x\right)  -\left(
E-V\left(  x\right)  \right)  R\left(  x\right)  =0 \label{4.5}%
\end{equation}
where $S\prime=v$ is the hydrodynamical flow velocity of the particle. We
assume that the solution $R$ takes the form%
\begin{equation}
R\left(  x\right)  =\sum_{j=0}^{\infty}\alpha_{j}x^{j}\cdot h\left(  x\right)
, \label{dr0}%
\end{equation}
for some constant $\alpha_{j}$ and a twice-differentiable function, $h$, that
represents the behavior of $\rho$ about its tails. Then, we have%

\begin{equation}
\frac{d}{dx}R\left(  x\right)  =\sum_{j=0}^{\infty}j\alpha_{j}x^{j-1}h\left(
x\right)  +\sum_{j=0}^{\infty}\alpha_{j}x^{j}h^{\prime}\left(  x\right)  ,
\label{dr1}%
\end{equation}
and%
\begin{equation}
\frac{d^{2}}{dx^{2}}R\left(  x\right)  =\sum_{j=0}^{\infty}\left(  j+1\right)
\left(  j+2\right)  \alpha_{j+2}x^{j}h\left(  x\right)  +\sum_{j=0}^{\infty
}\alpha_{j}x^{j}h^{\prime\prime}\left(  x\right)  +2\sum_{j=0}^{\infty}%
j\alpha_{j}x^{j-1}h^{\prime}\left(  x\right)  . \label{dr2}%
\end{equation}
Substituting (\ref{dr0}), (\ref{dr1}) and (\ref{dr2}) in (\ref{4.5}), we get%
\[
\sum_{j=0}^{\infty}\left[  \left(  j+1\right)  \left(  j+2\right)
\alpha_{j+2}h\left(  x\right)  +\alpha_{j}h^{\prime\prime}\left(  x\right)
+2j\alpha_{j}h^{\prime}\left(  x\right)  /x+\left(  E-V\left(  x\right)
-S^{\prime2}\right)  \alpha_{j}h\left(  x\right)  \right]  x^{j}=0.
\]
We decompose the constant part and the part that depends on $x,$ for each of
the components in the coefficients of $x^{j},$%
\begin{align*}
h\left(  x\right)   &  =C^{\left(  0\right)  }+Q^{\left(  0\right)  }\left(
x\right)  ,\text{ }h^{\prime}\left(  x\right)  /x=C^{\left(  1\right)
}+Q^{\left(  1\right)  }\left(  x\right)  ,\\
h^{\prime\prime}\left(  x\right)   &  =C^{\left(  2\right)  }+Q^{\left(
2\right)  }\left(  x\right)  ,\text{ }V\left(  x\right)  +S^{\prime
2}=C^{\left(  4\right)  }+Q^{\left(  4\right)  }\left(  x\right)  .
\end{align*}
In order to satisfy equation (\ref{4.5}), the terms that depend on $x$ have to
be equal to some real constant $b_{j}$. So%
\[
\sum_{j=0}^{\infty}\left[  \left(  j+1\right)  \left(  j+2\right)
\alpha_{j+2}C^{\left(  0\right)  }+\alpha_{j}C^{\left(  2\right)  }%
+2j\alpha_{j}C^{\left(  1\right)  }-\alpha_{j}\left(  E-C^{\left(  4\right)
}\right)  C^{\left(  0\right)  }+b_{j}\right]  x^{j}=0
\]
Now, for achieving the above equality, the coefficients of $x^{j}$ should be
zero. Then, after some algebraic calculation, we have%
\begin{equation}
\frac{\alpha_{j+2}}{\alpha_{j}}=-\frac{C^{\left(  2\right)  }+2jC^{\left(
1\right)  }-\left(  E-C^{\left(  4\right)  }\right)  C^{\left(  0\right)
}+b_{j}/\alpha_{j}}{\left(  j+1\right)  \left(  j+2\right)  C^{\left(
0\right)  }}. \label{EE3}%
\end{equation}
Since $\alpha_{j+2}/\alpha_{j}$ implies, in general, on a divergence sum
(\ref{dr0}), for having convergence, we impose the condition that the
nominator of the RHS of (\ref{EE3}) is zero. This produces our\ energy levels%
\begin{equation}
E=C^{\left(  4\right)  }+\frac{1}{C^{\left(  0\right)  }}\left(  C^{\left(
2\right)  }+\frac{b_{j}}{\alpha_{j}}+2C^{\left(  1\right)  }\cdot j\right)
,\text{ }j=0,1,2,...\text{ }. \label{EE4}%
\end{equation}

The proposed approach is not unique for the $1D$ case, and in fact, it holds
for a particle in arbitrary $N$ spatial dimensions $\boldsymbol{x=}\left(
x_{1},x_{2},...,x_{N}\right)  ^{T}$\bigskip.

\subsection{Quantization of the angular momentum}

The Schr\"{o}dinger equation shows that angular momentum takes discrete
values, i.e., it is quantized. But do the Madelung equations describe
quantized angular momentum? In the following, we show that by paying attention
to the properties of the probability density function $\rho,$ the quantization
of the angular momentum naturally appears in the proposed local geometrical
formalism of quantum mechanics. Let us consider the angular momentum operator
$\widehat{L}^{2}\ $in the polar coordinates $\left(  r,\theta,\phi\right)  \in%
\mathbb{R}
_{\geq}\times\left[  0,\pi\right]  \times\left[  0,2\pi\right]  .$ We consider
the separation of variables $\sqrt{\rho\left(  r,\theta,\phi\right)
}=R\left(  r\right)  \cdot\mathcal{R}\left(  \theta,\phi\right)
,\ $and\ consider the angular momentum operator in the angle coordinates
$\left(  \theta,\phi\right)  .$ We start with $\widehat{L}^{2}=-\left[
\frac{1}{\sin\left(  \theta\right)  }\frac{\partial}{\partial\theta}\left(
\sin\theta\frac{\partial}{\partial\theta}\right)  +\frac{1}{\sin^{2}\left(
\theta\right)  }\frac{\partial^{2}}{\partial\phi^{2}}\right]  ,$ and obtain
the real and imaginary parts%
\begin{equation}
\sin\left(  \theta\right)  \left(  \cos\theta\mathcal{R}_{\theta}+\sin\left(
\theta\right)  \left(  \mathcal{R}_{\theta\theta}-S_{\theta}^{2}%
\mathcal{R}\right)  \right)  +\left(  \mathcal{R}_{\phi\phi}-S_{\phi}%
^{2}\mathcal{R}\right)  +\lambda\sin^{2}\left(  \theta\right)  \mathcal{R}%
=0.\label{EQ_11}%
\end{equation}
and%
\begin{equation}
\sin\left(  \theta\right)  \left(  \cos\theta S_{\theta}+\sin\theta\left(
S_{\theta\theta}\mathcal{R}+2S_{\theta}\mathcal{R}_{\theta}\right)  \right)
+S_{\phi\phi}\mathcal{R}+2S_{\phi}\mathcal{R}_{\phi}=0,\label{EQ_22}%
\end{equation}
where $f_{x}:=df/dx,$ $f_{xy}:=d^{2}f/dxdy.$ Let us focus now on the real part
(\ref{EQ_11}). By taking $u=\cos\left(  \theta\right)  ,-1\leq u\leq1,$ eq.
(\ref{EQ_11}) takes the form%
\begin{equation}
\sin\left(  \theta\right)  \left(  -2\sin\left(  \theta\right)  \cos
\theta\mathcal{R}_{u}+\sin\left(  \theta\right)  \left(  \sin^{2}\left(
\theta\right)  \mathcal{R}_{uu}-S_{\theta}^{2}\mathcal{R}\right)  \right)
+\left(  \mathcal{R}_{\phi\phi}-S_{\phi}^{2}\mathcal{R}\right)  +\lambda
\sin^{2}\left(  \theta\right)  \mathcal{R}=0.\label{SSS1}%
\end{equation}
Dividing\ by $\sin^{2}\left(  \theta\right)  ,$ we have
\[
-2u\mathcal{R}_{u}+\left(  1-u^{2}\right)  \mathcal{R}_{uu}-S_{\theta}%
^{2}\mathcal{R}+\frac{1}{1-u^{2}}\left(  \mathcal{R}_{\phi\phi}-S_{\phi}%
^{2}\mathcal{R}\right)  +\lambda\mathcal{R}=0.
\]
We now consider a power sum for $\mathcal{R},$ $\mathcal{R}=\sum
_{j,k=0}^{\infty}a_{jk}u^{j}\phi^{k}.$ Then, we have%
\begin{align*}
&  \sum_{j,k=0}^{\infty}[\left(  j+1\right)  \left(  j+2\right)  a_{\left(
j+2\right)  k}-j\left(  j+1\right)  a_{jk}-S_{\theta}^{2}\mathcal{R}\\
&  +\frac{1}{1-u^{2}}\left(  \left(  k+1\right)  \left(  k+2\right)
a_{j\left(  k+2\right)  }-S_{\phi}^{2}a_{jk}\right)  +\lambda a_{jk}]u^{j}%
\phi^{k}\\
&  =0.
\end{align*}
For this equation to hold, we assume that all the parts in the coefficient of
$u^{j}\phi^{k}$ that are not constants\ cancel each other. We thus have%
\[
\sum_{j,k=0}^{\infty}[\left(  j+1\right)  \left(  j+2\right)  a_{\left(
j+2\right)  k}-j\left(  j+1\right)  a_{jk}+\lambda a_{jk}]u^{j}\phi^{k}=0.
\]
Then, for this equation to hold, the coefficients should sum up to zero,%
\[
\left(  j+1\right)  \left(  j+2\right)  a_{\left(  j+2\right)  k}-j\left(
j+1\right)  a_{jk}+\lambda a_{jk}=0,
\]
and after some calculations, we obtain%
\begin{equation}
\frac{a_{\left(  j+2\right)  k}}{a_{jk}}=-\frac{-j\left(  j+1\right)
+\lambda}{\left(  j+1\right)  \left(  j+2\right)  }.\label{Eq3}%
\end{equation}
Now, in the limit $j\rightarrow\infty$ we get, in general, a divergence of the
density function. Thus, we should only consider a finite power sum of
$\mathcal{R}$ with respect to $u,$ and so we finally obtain the quantized
angular momentum by taking the numerator at the RHS\ to be zero, and we get
$\lambda=j\left(  j+1\right)  $. Following the angular momentum operator
$\widehat{L}^{2}=-\left(  r^{2}\frac{d^{2}}{dr^{2}}-\frac{d}{dr}r^{2}\frac
{d}{dr}\right)  ,$ the\ eigenvalue equation gives us the\ real and imaginary
parts of the Madelung equations as\ a system of decoupled equations%
\begin{equation}
2rR\left(  r\right)  ^{\prime}=\lambda R\left(  r\right)  \Longrightarrow
R\left(  r\right)  =c_{1}r^{\lambda/2},\label{EQ1}%
\end{equation}%
\begin{equation}
rS\left(  r\right)  ^{\prime}=0.\label{EQ2}%
\end{equation}
Substituting $\lambda=j\left(  j+1\right)  ,$ the first equation gives the
radial part of the quantized angular momentum, $R\left(  r\right)
=c_{1}r^{j\left(  j+1\right)  /2}.$ We note that in the cylindrical coordinate
system $\left(  z,\rho,\phi\right)  $ it seems, at first glance, that we have
to use an external quantum condition for getting quantized angular momentum.
However, as we observe that $L_{z}$ is still a component of the angular
momentum, it must have integer values following a similar approach given above
for the polar coordinate system. Another point we would like to emphasize is
that\ in two dimensions, one cannot obtain quantization using the Madelung
equations only and must have an external quantization condition. For example,
suppose we have a central potential in the form of a harmonic oscillator
$V\left(  r\right)  =\frac{1}{2}\kappa r^{2},$ with the underlying polar
coordinates $\left(  r,\phi\right)  .$ Then, it is straightforward that the
solutions of the Schr\"{o}dinger equation\ $R\left(  r\right)  e^{im\phi}$
cannot be quantized in the representation of $\rho$ and $J,$
i.e.,\ non-integer $m$ is still a valid solution in the picture of the
Madelung equations.

\end{document}